\documentstyle[aps,eqsecnum]{revtex}
\begin{document}
\draft 
\title{Implications of invariance of the Hamiltonian under canonical 
transformations in phase space}
\author{E. D. Davis and G. I. Ghandour}
\address{Physics Department, Kuwait University, P. O. Box 5969, Safat, Kuwait}
\date{April, 1999}
\maketitle 
\begin{abstract}
We observe that, within the effective generating function formalism for the 
implementation of canonical transformations within wave mechanics,
non-trivial canonical transformations which leave invariant the form of the 
Hamilton function of the classical analogue of a quantum system manifest 
themselves in an integral equation for its stationary state eigenfunctions. We 
restrict ourselves to that subclass of these dynamical symmetries for which 
the corresponding effective generating functions are necessarily free of 
quantum corrections. We demonstrate that infinite families of such 
transformations exist for a variety of familiar conservative systems of one 
degree of freedom. We show how the geometry of the canonical transformations 
and the symmetry of the effective generating function can be exploited to pin 
down the precise form of the integral equations for stationary state 
eigenfunctions. We recover several integral equations found in the literature 
on standard special functions of mathematical physics. We end with a brief 
discussion (relevant to string theory) of the generalization to scalar field 
theories in 1+1 dimensions. 
\end{abstract}

\section{Introduction}

Given the formal similarities between quantum mechanics and the Hamiltonian 
formulation of classical mechanics, it is not surprizing that there have been 
several attempts to define within quantum mechanics transformations analogous 
to the canonical transformations underpinning the powerful Hamilton-Jacobi 
method. The founding fathers of quantum mechanics were content to identify the 
quantum analogues of canonical transformations as unitary transformations of 
the position and momentum operators which preserve the canonical commutation 
relations~\cite{BHJ26,Di26}. However, to enhance the scope
of applications, various operator-based extensions of this notion of a quantum 
canonical transformation have been proposed~\cite{An94,LY94}. The path-integral
formulation of quantum mechanics (and quantum field theory) with its c-number 
representation of co-ordinates and momenta has also prompted more ambitious 
constructions~\cite{Ka84}, but these are not without their 
complications~\cite{BKN90} (because of the ambiguous interrelationships between 
the momenta $p_i$ and co-ordinates $q_i$ introduced in the discretization of 
the path integrals~\cite{Ga66}). Finally, there are the realizations of
canonical transformations within quantum theory as integral 
transforms~\cite{Hi82,Gh87}.

It has been observed~\cite{Gh87} that, at least for time-independent problems, 
there is a simple implementation of canonical transformations within wave 
mechanics via an integral transform of the form 
\begin{equation}\label{eq:qctransf}
\phi_\alpha(q)=n_\alpha\int e^{(i/\hbar)F(q,Q)}\Phi_\alpha(Q){\rm d}Q.
\end{equation}
Here, $\phi_\alpha(q)$ and $\Phi_\alpha(Q)$ denote the wavefunctions of the 
same state $\left|\alpha\right\rangle$ for two choices of generalized 
co-ordinate $q$ and $Q$, $n_\alpha$ is a state-dependent relative 
normalization (independent of $q$ and $Q$) and, to leading order in $\hbar$, 
the 
state-independent quantum (or {\em effective}) generating function $F(q,Q)$ 
coincides with the generating function ${\cal F}(q,Q)$ of the canonical 
transformation $q,p\longrightarrow Q,P$ within classical mechanics. As we 
demonstrate below (in Sec.~\ref{sc:QCT}), this relation between classical and
quantum generating functions respects the algebra of composition of canonical 
transformations and, under certain broad conditions (spelt out in 
Sec.~\ref{sc:QCT}), the quantum corrections to ${\cal F}(q,Q)$ vanish. 
(Unless stated otherwise, we assume throughout this work that we are dealing
with systems of one degree of freedom but use of the quantum 
canonical transform in Eq.~(\ref{eq:qctransf}) is not limited to such 
systems~\cite{CZ94,Lo95,Bo97,EG99}.) 

Quantum realizations of canonical transformations provide elegant solutions to 
eigenvalue problems in quantum mechanics~\cite{EG99,An95,Yi96}. They also offer
a platform for the exploration of a novel kind of issue, namely the impact at 
the quantum level of symmetries which hold in the {\em full\/} phase space at 
the classical level. A topical example would be the investigation of duality in 
field theories beyond the classical level~\cite{Lo95,Sf98}. In this paper, we
consider the implications of the existence of non-trivial canonical 
transformations $q,p\longrightarrow Q,P$ in the full phase space such that the 
transformed Hamiltonian function (or Kamiltonian) ${\cal K}(Q,P)\equiv {\cal H}
(q(Q,P),p(Q,P))$ is the same as the original Hamiltonian function ${\cal H}$, 
i.e.\ ${\cal K}(Q,P)={\cal H}(Q,P)$. To this end, we adopt the effective 
generating function formalism of~\cite{Gh87}.

Canonical transformations which preserve the form of a Hamilton function $\cal 
H$ have a dynamical significance within the classical mechanics of systems of 
one degree of freedom: they map trajectories of such systems (level curves of 
$\cal H$) onto themselves and, thus, amount to evolutions in time. Within wave
mechanics, we have for a form-preserving transformation $q,p\longrightarrow 
Q,P$ the equality $\Phi_\alpha(Q)=\phi_\alpha(Q)$ (because the Hamiltonian 
operators for $q$ and $Q$ are identical in form) and the integral transform in 
Eq.~(\ref{eq:qctransf}) reduces to an integral equation for stationary state 
eigenfunctions.

The point of departure for the present study is the conjecture that the class 
of (classical) generating functions ${\cal F}(q,Q)$ of form-preserving 
canonical transformations may include members for which quantum corrections 
vanish. We restrict ourselves to Hamiltonians describing a particle of 
mass $m$ in the potential ${\cal V}(q)$. We demonstrate (in Sec.~\ref{sc:ihf} 
below) that, in addition to the free theory (${\cal V}\equiv 0$), there are 
six distinct choices of potential ${\cal V}(q)$ for which infinite families 
$\{{\cal F}_\mu(q,Q)\}$ of non-trivial form-preserving and correction-free 
generating functions exist ($\mu$ labels the members of these families).
These potentials include some of the most ubiquitous (notably the linear and 
quadratic potentials) and the integral equations implied 
\begin{equation}\label{eq:bie}
\psi_\alpha(q)={\cal N}_\alpha(\mu)\int e^{(i/\hbar){\cal F}_\mu(q,Q)}
\psi_\alpha(Q){\rm d}Q
\end{equation}
apply to several standard special functions of mathematical physics: in the 
notation of~\cite{AS64}, the parabolic cylinder functions $D_n$ (integer order), 
the Airy function $\mbox{Ai}$, the modified Bessel functions $K_\sigma$ 
(imaginary order), the Mathieu functions $ce_r$ and $se_r$ and the modified 
Mathieu functions $Mc_r^{(1)}$ and $Ms_r^{(1)}$.

The major preoccupation of this paper is to show that not only can the kernel 
$e^{(i/\hbar){\cal F}_\mu(q,Q)}$ in these integral equations be obtained via 
consideration of canonical transformations, but so can the reciprocals 
${\cal N}_\alpha(\mu)$ of its eigenvalues (modulo, in some cases, a phase). Two 
distinct and complementary methods apply. For those potentials for which the 
canonical transformations form an abelian group (the linear and quadratic 
potentials), the composition of canonical transformations implies a functional 
relation for ${\cal N}_\alpha(\mu)$ which determines it up to a phase. For the 
other potentials, we can take advantage of a remarkable symmetry in the 
dependence of the corresponding generating functions ${\cal F}_\mu(q,Q)$ on 
$\mu$, $q$ and $Q$ (described in Sec.~\ref{sc:op}).

We close in Sec.~\ref{sc:dis} with some remarks on lines of investigation 
suggested by this work.

\section{The quantum canonical transform}\label{sc:QCT}

Let $\{\phi_\alpha(q)\}$ denote the complete set of stationary wavefunctions of 
a quantum system with Hamiltonian operator $\widehat{h}\left(q,
{\textstyle\frac{\hbar}{i}\frac{\partial\ }{\partial q}}\right)$ and let 
$\widehat{H}\left(Q,{\textstyle\frac{\hbar}{i}\frac{\partial\ }{\partial Q}}
\right)$ be the realization of the Hamiltonian of the system for another choice 
of generalized co-ordinate $Q$ and $\{\Phi_\alpha(Q)\}$ the corresponding 
complete set of stationary wavefunctions. The quantum canonical transform 
introduced in~\cite{Gh87} relates the $\phi_\alpha(q)$'s to the 
$\Phi_\alpha(Q)$'s via an integral relationship of the form in 
Eq.~(\ref{eq:qctransf}). The quantum generating function $F(q,Q)$ is fixed by 
the requirement that $\phi_\alpha(q)$ and $\Phi_\alpha(Q)$ are eigenfunctions 
of the same complete set of commuting observables with exactly the same set of 
quantum numbers $\alpha$. The state-dependent relative normalization $n_\alpha$
is chosen so that the normalizations of $\phi_\alpha(q)$ and $\Phi_\alpha(Q)$ 
are compatible. 

For a system with a non-degenerate energy spectrum (like those considered in 
Secs.~\ref{sc:qlp} and~\ref{sc:op} below), the restriction on $F(q,Q)$ reduces 
to the condition that $\phi_\alpha(q)$ and $\Phi_\alpha(Q)$ are eigenfunctions 
of $\widehat{h}$ and $\widehat{H}$, respectively, with the same energy 
$E_\alpha$. Substituting for $\phi_\alpha$ in $\widehat{h}\phi_\alpha=E_\alpha
\phi_\alpha$ using Eq.~(\ref{eq:qctransf}) and then replacing the product
$E_\alpha\Phi_\alpha$ by $\widehat{H}\Phi_\alpha$, we find that, after the 
requisite number of integration by parts and appealing to the completeness of 
the $\Phi_\alpha$'s, this condition implies $F(q,Q)$ should satisfy
\begin{equation}\label{eq:Feqn}
\widehat{h}\left(q,{\textstyle\frac{\hbar}{i}\frac{\partial\ }{\partial q}}
\right)e^{(i/\hbar)F(q,Q)}=
\widehat{H}\left(Q,-{\textstyle\frac{\hbar}{i}\frac{\partial\ }{\partial Q}}
\right)e^{(i/\hbar)F(q,Q)}
\end{equation}
provided the endpoint terms generated in the integration by parts vanish. These 
terms take the form of the bilinear combination
\begin{equation}\label{eq:bilcom}
e^{(i/\hbar)F(q,Q)}\frac{\partial\ }{\partial Q}\Phi_\alpha-\Phi_\alpha\frac{
\partial\ }{\partial Q}e^{(i/\hbar)F(q,Q)}
\end{equation}
for the Hamiltonian
\begin{equation}\label{eq:H}
\widehat{H}\left(Q,{\textstyle\frac{\hbar}{i}\frac{\partial\ }{\partial Q}}
\right)=-\frac{\hbar^2}{2m}\frac{\partial^2\ }{\partial Q^2} + V(Q)
\end{equation}
considered below. For bound states, the vanishing of
the wavefunction and its first derivative at infinity guarantees that
the bilinear concomitant in Eq.~(\ref{eq:bilcom}) is zero.

The relation of the quantum generating function to a generating function 
$f_{\rm cl}(q,Q)$ of a canonical transformation within the Hamiltonian 
formulation of classical mechanics can be brought out by adopting for $F(q,Q)$ 
an expansion in powers of $i\hbar$:
\[F(q,Q)=\sum\limits_{n=0}^\infty{\cal F}_n(q,Q)(i\hbar)^n.\]
If we take $\widehat{H}\left(Q,{\textstyle\frac{\hbar}{i}\frac{\partial\ }{
\partial Q}}\right)$ to be given by Eq.~(\ref{eq:H}) and $\widehat{h}
\left(q,{\textstyle\frac{\hbar}{i}\frac{\partial\ }{\partial q}}\right)$ to be 
given by
\[\widehat{h}\left(q,{\textstyle\frac{\hbar}{i}\frac{\partial\ }{\partial q}}
\right)=-\frac{\hbar^2}{2m}\frac{\partial^2\ }{\partial q^2} + v(q),\]
then substitution of this expansion into Eq.~(\ref{eq:Feqn}) yields for ${\cal 
F}_0$
\begin{equation}\label{eq:f0eqn}
\frac{1}{2m}\left(\frac{\partial{\cal F}_0}{\partial q}\right)^2 + v(q)
=\frac{1}{2m}\left(-\frac{\partial{\cal F}_0}{\partial Q}\right)^2 + V(Q)
\end{equation}
and for the other ${\cal F}_n$'s ($n>0$)
\begin{equation}\label{eq:fneqn}
\sum\limits_{k=0}^n\left(\frac{\partial{\cal F}_k}{\partial q}\frac{\partial{
\cal F}_{n-k}}{\partial q}-\frac{\partial{\cal F}_k}{\partial Q}\frac{\partial{
\cal F}_{n-k}}{\partial Q}\right)=\frac{\partial^2{\cal F}_{n-1}}{\partial 
q^2}-\frac{\partial^2{\cal F}_{n-1}}{\partial Q^2}.
\end{equation}
Equation (\ref{eq:f0eqn}) is automatically satisfied if we identify
${\cal F}_0(q,Q)$ as the classical generating function of a canonical 
transformation $(q,p)\longrightarrow(Q,P)$ for which the original Hamiltonian 
function  ${\cal H}(q,p)=p^2/(2m)+v(q)$ and the transformed Hamiltonian 
function (or Kamiltonian) ${\cal K}(Q,P) =P^2/(2m)+V(Q)$. 
If $\partial^2{\cal F}_0/\partial q^2=\partial^2{\cal F}_0/
\partial Q^2$, i.e.\ ${\cal F}_0$ is of the form
\begin{equation}\label{eq:dAlembert}
{\cal F}_0(q,Q)={\cal F}_+(q_+)+{\cal F}_-(q_-),
\end{equation}
where $q_\pm\equiv(q\pm Q)/2$ and ${\cal F}_+$ and ${\cal F}_-$ are arbitrary 
functions, then Eq.~(\ref{eq:fneqn}) implies that the quantum corrections 
${\cal F}_n$ ($n\ge 1$) can be taken to be zero.

The parallel between quantum and classical generating functions also extends to 
the composition of transformations. Let $F_1(q,q_i)$ and $F_2(q_i,Q)$ denote 
the quantum generating functions for the canonical transformations 
$(q,p)\longrightarrow(q_i,p_i)$ and $(q_i,p_i)\longrightarrow (Q,P)$, 
respectively. The exact relation among these generating functions and the 
quantum generating function $F_c(q,Q)$ for the composition $(q,p)
\longrightarrow(Q,P)$ reads
\begin{equation}\label{eq:qgfconv}
n_\alpha^{(1)}n_\alpha^{(2)}\int e^{(i/\hbar)[F_1(q,q_i)+F_2(q_i,Q)]}{\rm d}q_i
=n_\alpha^{(c)}e^{(i/\hbar)F_c(q,Q)}.
\end{equation}
To identify the relation between the classical contribution $f_c$ to $F_c$
and the classical contributions to $F_1$ and $F_2$ (assumed to be $f_1$ and 
$f_2$, respectively), we can evaluate the integration over $q_i$ in 
Eq.~(\ref{eq:qgfconv}) in the stationary phase approximation. Retaining only 
the terms most singular in $\hbar$, we obtain
\begin{equation}\label{eq:qconvapp}
n_\alpha^{(c)}e^{(i/\hbar)f_c(q,Q)}=\sqrt{\frac{2\pi\hbar i}{\kappa}}
n_\alpha^{(1)}n_\alpha^{(2)}e^{(i/\hbar)f_s(q,Q)},
\end{equation}
where
\begin{equation}\label{eq:cgfconv}
f_s(q,Q)\equiv f_1(q,\overline{q_i})+f_2(\overline{q_i},Q)
\end{equation}
with $\overline{q_i}$ chosen so that
\begin{equation}\label{eq:stationary}
\left.\frac{\partial\ }{\partial q_i}
\left(f_1(q,q_i)+f_2(q_i,Q)\right)\right|_{q_i=\overline{q_i}}=0
\end{equation} 
and we have made the generically valid assumption that
\[\kappa\equiv\left.\left(\frac{\partial^2f_1}{\partial q_i^2}
+\frac{\partial^2f_2}{\partial q_i^2}\right)\right|_{q_i=\overline{q_i}}\]
is non-zero. (For simplicity, we have also assumed that there is only one 
stationary point.) Interpretation of Eq.~(\ref{eq:qconvapp}) is complicated by 
the unknown, but, in general, singular dependence of the relative 
normalizations $n_\alpha^{(1)}$, $n_\alpha^{(2)}$ and $n_\alpha^{(c)}$ on 
$\hbar$. Nevertheless, the essential singularities in $\hbar$ in 
Eq.~(\ref{eq:qconvapp}) can only match if the dependence on $q$ and $Q$ 
cancels --- i.e.\ $f_c$ and $f_s$ differ at most by a constant. Exactly the 
same relation between $f_c$ and $f_s$ is implied by the canonical formalism of 
classical mechanics. We have that
\begin{eqnarray*}
\frac{\partial f_s}{\partial q}&=&\frac{\partial f_1}{\partial q}+\left(
\frac{\partial f_1}{\partial q_i}+\frac{\partial f_2}{\partial q_i}\right)_{
|_{q_i=\overline{q_i}}}\frac{\partial\overline{q_i}}{\partial q}=
\frac{\partial f_1}{\partial q}=+p,\\
\frac{\partial f_s}{\partial Q}&=&\frac{\partial f_2}{\partial Q}+
\left(\frac{\partial f_1}{\partial q_i}+\frac{\partial f_2}{\partial q_i}
\right)_{|_{q_i=\overline{q_i}}}\frac{\partial\overline{q_i}}{\partial Q}=
\frac{\partial f_2}{\partial Q}=-P,
\end{eqnarray*}
i.e.\ the partial derivatives of $f_s$ coincide with those of $f_c$.

Further parallels with generating functions in the canonical formalism of 
classical mechanics are discussed in~\cite{Ma88,EG99}. 

\section{Invariant Hamiltonian functions}\label{sc:ihf}

Quantum generating functions $F(q,Q)$ which give rise to integral equations
can be obtained by determining {\em classical\/} generating functions 
${\cal F}_0(q,Q)$ of the form in Eq.~(\ref{eq:dAlembert})
which induce canonical transformations $q,p\longrightarrow Q,P$
such that the Kamiltonian function ${\cal K}(Q,P)\equiv {\cal H}(q(Q,P),
p(Q,P))$ is the same as the original Hamiltonian function ${\cal H}$, i.e.\ 
${\cal K}(Q,P)={\cal H}(Q,P)$. (Below, we drop the
subscript $0$, denoting ${\cal F}_0$ by $\cal F$.)

We consider a canonically conjugate pair of variables $q$ and $p$ for which the 
Hamiltonian function 
\[{\cal H}(q,p)=\frac{p^2}{2m}+{\cal V}(q).\]
Inspection of the relations for the momenta in terms of a classical generating 
function ${\cal F}(q,Q)$
\begin{equation}\label{eq:mom}
p=\frac{\partial{\cal F}}{\partial q}=\frac{1}{2}
[{\cal F}_-^{\,\prime}(q_-)+{\cal F}_+^{\,\prime}(q_+)]\qquad\qquad
P=-\frac{\partial{\cal F}}{\partial Q}=\frac{1}{2}
[{\cal F}_-^{\,\prime}(q_-)-{\cal F}_+^{\,\prime}(q_+)]
\end{equation}
shows that at least for the case of the free theory ($V\equiv 0$)
there are non-trivial canonical transformations under which
the form of the Hamiltonian is unchanged: for the generating functions 
${\cal F}_{\rm free}^{\,+}={\cal F}_-(q_-)$ and ${\cal F}_{\rm free}^{\,-}=
{\cal F}_+(q_+)$, where ${\cal F}_+$ and ${\cal F}_-$ are arbitrary, the 
transformed generalized momentum $P=+p$ and $P=-p$, respectively, from which 
the form invariance of the free Hamiltonian ${\cal H}_{\rm free}=p^2/(2m)$ is 
obvious. 

More generally, substituting for the momenta in  
\[\frac{p^2}{2m}+{\cal V}(q)=\frac{P^2}{2m}+{\cal V}(Q)\]
using Eq.~(\ref{eq:mom}), we deduce that the following relation must hold 
between $\cal V$ and ${\cal F}_\pm$:
\begin{equation}\label{eq:basic}
\frac{1}{2m}{\cal F}_+^{\,\prime}(q_+){\cal F}_-^{\,\prime}(q_-)={\cal V}(q_+ 
-q_-)-{\cal V}(q_+ +q_-).
\end{equation}
To proceed, we assume that ${\cal F}_-^{\,\prime}$ and $\cal V$ are analytic 
and we expand both sides of Eq.~(\ref{eq:basic}) in powers of $q_-$ to obtain 
($k=0,1,2,\ldots$)
\begin{equation}\label{eq:ebasic}
\frac{1}{2m}{\cal F}_+^{\,\prime}(x){\cal F}_-^{(k+1)}(0)= [(-1)^k-1]{\cal 
V}^{(k)}(x),
\end{equation}
where we have set $q_+=x$. If we ignore the possibility that ${\cal V}^\prime(
x)\equiv 0$ (in which case we recover the results given above for the free 
theory), then Eq.~(\ref{eq:ebasic}) for $k=1$ implies that $\mu\equiv {\cal 
F}_-^{(2)}(0)\ne 0$ and
\begin{equation}\label{eq:fprime}
{\cal F}_+^{\,\prime}(x)=-\frac{4m}{\mu}{\cal V}^\prime(x).
\end{equation}
Substituting Eq.~(\ref{eq:fprime}) into Eq.~(\ref{eq:ebasic}), it reduces to
the simultaneous requirements that {\em odd\/} derivatives of $\cal V$ are 
given by
\begin{equation}\label{eq:Vbasic}
{\cal V}^{(k)}(x)=\frac{{\cal F}_-^{(k+1)}(0)}{{\cal F}_-^{(2)}(0)}{\cal 
V}^\prime(x)
\end{equation}
and that ${\cal F}_-$ is even (so that the derivatives ${\cal F}_-^{(n)}(0)=0$ 
for $n$ odd). 

From Eq.~(\ref{eq:fprime}), we immediately have that
\[{\cal F}_+(x)=-\frac{4m}{\mu}{\cal V}(x),\]
where we have dropped an irrelevant constant of integration.
The implications of Eq.~(\ref{eq:Vbasic}) for ${\cal V}$ and ${\cal F}_-$ 
depend on whether or not the third derivative ${\cal V}^{(3)}$ vanishes. 

If ${\cal V}^{(3)}(x)\ne 0$, then Eq.~(\ref{eq:Vbasic}) implies that $\rho
\equiv{\cal F}_-^{(4)}(0)/{\cal F}_-^{(2)}(0)\ne 0$, ${\cal V}^{(3)}(x)=\rho 
V^{(1)}(x)$ and $g^{(2k)}(0)=\rho^{k-1}\mu$ for $k>1$. Thus, since $\rho$ may 
be of either sign, the potential can either be the combination of hyperbolic 
functions ($\rho=+\beta^2>0$)
\[{\cal V}_+(x)=A\cosh\beta x+B\sinh\beta x\]
or the combination of sinusoidal functions ($\rho=-\beta^2<0$)
\[{\cal V}_-(x)=A\cos\beta x+B\sin\beta x,\]  
where $\beta$, $A$ and $B$ are arbitrary constants. The corresponding forms of 
${\cal F}_-$ are (up to arbitrary additive constants)
\[{\cal F}_-^{\,+}(x)=\frac{\mu}{\beta^2}\cosh\beta x\]
and
\[{\cal F}_-^{\,-}(x)=-\frac{\mu}{\beta^2}\cos\beta x,\]
respectively. If we assume that ${\cal V}^{(3)}(x)\equiv 0$, then 
Eq.~(\ref{eq:Vbasic}) implies that, with the exception of ${\cal F}_-^{(2)}(0)$,
all the even derivatives ${\cal F}_-^{(2n)}(0)=0$. Discarding arbitrary 
additive constants, the potential is of the quadratic form
${\cal V}(x)=Ax^2+Bx$, where, as above, $A$ and $B$ are arbitrary constants,
and ${\cal F}_-$ is the quadratic ${\cal F}_-(x)=\mu x^2/2$.

Not only are there several classes of potential ${\cal V}(q)$ compatible with 
Eq.~(\ref{eq:basic}), but also, for each class of potentials, there is an 
infinite family of canonical transformations distinguished by different values 
of the parameter $\mu$ which is not fixed by the considerations above.
By translation of the origin or the inversion $q\longrightarrow -q$ or 
translation followed by inversion, all members of the classes of non-trivial 
potentials identified above can be reduced to one of those in 
Table~\ref{tb:potentials}. We also include the families of generating functions 
${\cal F}_\mu$ of canonical transformations which leave the corresponding 
Hamiltonians unchanged. In what follows, we take $\lambda>0$, a choice
which embraces the physically more interesting scenarios.

Below [in Sec.~(\ref{sc:op})], we shall have cause to invoke the 
limit $\mu\rightarrow\infty$. From Eqs.~(\ref{eq:mom}) and (\ref{eq:fprime}),
the difference in momenta
\begin{equation}\label{eq:pdiff}
P-p=\frac{4m}{\mu}{\cal V}^\prime(q_+).
\end{equation}
The difference in the coordinates $Q-q$ ($=-2q_-$) in terms of $p+P$ 
can be obtained by inversion of the relation
\begin{equation}\label{eq:qdiff}
P+p={\cal F}_-^{\,\prime}(q_-)
\end{equation}
implied by Eq.~(\ref{eq:mom}). Together, Eqs.~(\ref{eq:pdiff}) and 
(\ref{eq:qdiff}) imply that, for our choices of ${\cal F}_-$, both
$P-p=$ and $Q-q$ are of order $1/\mu$ for large $\mu$. [In the case of the 
sinusoidal potential, use of the principal value of arcsin in the 
inversion of Eq.~(\ref{eq:qdiff}) is understood.] Hence, in the limit
$\mu\rightarrow\infty$, all the canonical transformations of interest
reduce to the identity transformation.

\section{Integral equations for the quadratic and linear 
potentials}\label{sc:qlp}

It is natural to ask whether any of the families of canonical transformations 
we have identified are groups? At the level of generating functions, this 
question translates into whether when two generating functions 
${\cal F}_{\mu_1}$ and ${\cal F}_{\mu_2}$ of a particular family in 
Table~\ref{tb:potentials} are combined according to Eq.~(\ref{eq:cgfconv}), the 
resulting generating function ${\cal F}_{\mu_1\mu_2}$ is also a member of that 
family (to within an additive constant). This is not automatic since these 
families of generating functions have been constructed by restricting their 
members to be of the special form in Eq.~(\ref{eq:dAlembert}) and this 
constraint cannot, in general, be respected by the prescription in 
Eq.~(\ref{eq:cgfconv}). However, it turns out that, because of the simple 
geometrical operations in phase space effected by members of the families of 
canonical transformations corresponding to the quadratic and linear potentials, 
these two families do form groups. We are able to exploit the abelian character 
of these groups to convert Eq.~(\ref{eq:qconvapp}) into a functional 
relationship for the reciprocals ${\cal N}_\alpha(\mu)$ of eigenvalues and so 
deduce the integral equations for the quadratic and linear potentials.

\subsection{The quadratic potential}

The canonical transformations which leave the Hamiltonian function associated 
with this potential unchanged are linear transformations of the phase plane 
and so must be as either a rotation or an area-preserving shear or an 
area-preserving squeeze~\cite{PR82}. The observation that, for the rescaled 
canonical variables $q_s\equiv(m\lambda)^{1/4}q$ and $p_s\equiv p/(m
\lambda)^{1/4}$, the Hamiltonian function  
\[{\cal H}=\frac{1}{2}\sqrt{\frac{\lambda}{m}}\left(p_s^2+q_s^2\right)\]
suggests that it should be possible to decompose these transformations in terms 
of a rotation as
\[\left[\begin{array}{c}Q\\ P\end{array}\right]=
\left(\begin{array}{cc}(m\lambda)^{-1/4}&0\\ 0&(m\lambda)^{1/4}\end{array}
\right)
\left(\begin{array}{cc}\cos\theta&\sin\theta\\ -\sin\theta&\cos\theta
\end{array}\right)
\left(\begin{array}{cc}(m\lambda)^{1/4}&0\\ 0&(m\lambda)^{-1/4}\end{array}
\right)\left[\begin{array}{c}q\\ p\end{array}\right],\]
where the angle of rotation $\theta$ is dependent on the choice $\mu$. In fact, 
this does prove to be the case with the free parameter $\mu$ uniquely related 
to the angle of rotation $\theta$ by
\[\mu=-2\sqrt{\lambda m}\cot\frac{\theta}{2}.\]
The totality of these linear transformations thus constitutes a faithful matrix 
representation of the rotation group SO(2).

It is convenient to work with generating functions parametrized by $\theta$ 
instead of $\mu$, namely
\begin{equation}\label{eq:Ftheta}
F(q,Q|\theta)=\frac{1}{2}m\omega\left[2\csc\theta qQ-\cot\theta(q^2+Q^2)\right],
\end{equation}
where $\omega\equiv\sqrt{\lambda/m}$.
We write the corresponding integral equation for eigenfunctions
$\{\psi_n(q)\}$ ($n$ a non-negative integer) of the harmonic oscillator 
hamiltonian
\[\widehat{h}_{\rm ho}=
-\frac{\hbar^2}{2m}\frac{\partial^2\ }{\partial q^2} + \frac{1}{2}m\omega^2q^2\]
as
\begin{equation}\label{eq:hoint}
\psi_n(q)=N_n(\theta)\int e^{(i/\hbar) F(q,Q|\theta)}\psi_n(Q)
{\rm d}Q.
\end{equation}
Since $\widehat{h}_{\rm ho}$ commutes with the parity operator $\widehat{P}$ 
and has a non-degenerate spectrum, its eigenfunctions are automatically either 
even or odd. Although 
the generating function $F(q,Q|\theta)$ has been constructed only
with a view to ensuring that the integral on the righthand-side
of Eq.~(\ref{eq:hoint}) is an eigenfunction of $\widehat{h}_{\rm ho}$ with
the same eigenenergy as $\psi_n(q)$,
the dependence of $F(q,Q|\theta)$ on $q$ and $Q$ guarantees that
this integral has also the same parity as $\psi_n(q)$. The generating
function pertinent to the sinusoidal potential shares this property.

Use of $\theta$ in Eq.~(\ref{eq:hoint}) simplifies the treatment of the 
composition of transformations. We can immediately identify the quantum 
generating function $F_c(q,Q)$ for the composition of two transformations with 
generating functions $F_1=F(q,q_i|\theta_1)$ and $F_2= F(q_i,Q|\theta_2)$, 
respectively, as $F_c(q,Q)=F(q,Q|\theta_1+\theta_2)$. Accordingly, in this 
context, Eq.~(\ref{eq:qgfconv}) reads 
\[N_n(\theta_1) N_n(\theta_2)\int e^{(i/\hbar)[F(q,q_i|\theta_1)+F(q_i,Q|
\theta_2)]}{\rm d}q_i= N_n(\theta_1+\theta_2)
e^{(i/\hbar)F(q,Q|\theta_1+\theta_2)},\]
which, on evaluation of the gaussian integral over $q_i$, becomes
\begin{equation}\label{eq:Nhofun}
N_n(\theta_1+\theta_2)=\sqrt{\frac{2\pi\hbar}{m\omega i}}
\sqrt{\frac{\sin\theta_1\sin\theta_2}{\sin(\theta_1+\theta_2)}}
N_n(\theta_1)N_n(\theta_2).
\end{equation}
The functional relation in Eq.~(\ref{eq:Nhofun}) has the solution
\begin{equation}\label{eq:Nhosol}
N_n(\theta)=\sqrt{\frac{m\omega i}{2\pi\hbar}}\frac{e^{c_n\theta}}{\sqrt{
\sin\theta}},
\end{equation}
where the coefficient $c_n$ is arbitrary.

The coefficient $c_n$ can be fixed by 
appealing to the fact that, under the inversion $q,p\longrightarrow -q,-p$
[corresponding to the choice of $\theta=\pi$ in $F(q,Q|\theta)$],
the eigenfunctions $\psi_n(q)$ transform in a well-defined manner:
$\psi_n(-q)=(-1)^n\psi_n(q)$. From Eqs.~(\ref{eq:Ftheta}) and (\ref{eq:Nhosol}),
\[\lim_{\theta\rightarrow\pi} N_n(\theta)e^{(i/\hbar) F(q,Q|\theta)}=ie^{
c_n\pi}\delta(q+Q), \]
which, on substitution in Eq.~(\ref{eq:hoint}), implies $\psi_n(q)=
ie^{c_n\pi}\psi_n(-q)$. Consistency with the parity properties of
the $\psi_n(q)$'s is achieved by taking $c_n=-(n+1/2)i$.

In its final form, our integral equation for the eigenfunctions $\psi_n(q)$ of 
the harmonic oscillator reads
\begin{equation}\label{eq:hofinal}
\psi_n(q)=\sqrt{\frac{m\omega i}{2\pi\hbar}}\frac{e^{-i(n+1/2)\theta}}{\sqrt{
\sin\theta}}\int e^{i[m\omega/(2\hbar)]\left[2\csc\theta qQ-\cot\theta(q^2+Q^2)
\right]}\psi_n(Q){\rm d}Q.
\end{equation} 
To make the connection with known results, we note that we may read off from
Eq.~(\ref{eq:hofinal}) that the expansion coefficients of $N_0(\theta)e^{(i/
\hbar)F(q,Q|\theta)}$ in the orthonormal basis $\{\psi_n(Q)\}$ are $e^{in
\theta}\psi_n(q)$ and, hence, construct the identity
\[e^{i[m\omega/(2\hbar)]\left[2\csc\theta qQ-\cot\theta(q^2+Q^2)\right]}=
\sqrt{\frac{2\pi\hbar}{m\omega i}}\sqrt{\sin\theta}
\sum_n e^{+i(n+1/2)\theta}\psi_n(q)\psi_n(Q).\]
Invoking the relation of the $\psi_n(q)$'s to the parabolic cylinder function 
$D_n(x)$~\cite{AS64}, namely
\[\psi_n(q)=\left(\frac{m\omega}{\pi\hbar}\right)^{1/4}\frac{1}{\sqrt{n!}}
D_n(\sqrt{2m\omega/\hbar}q),\]
we recover one of the addition theorems for the $D_n$'s given in chapter 11 
of~\cite{MF53}. (To obtain this addition theorem in the precise form given 
in~\cite{MF53}, we must set $e^{i\theta}=i\tan\phi$, $\sqrt{m\omega/\hbar}q=
e^{i\pi/4}\lambda$ and $\sqrt{m\omega/\hbar}Q=e^{i3\pi/4}\mu$.) 

\subsection{The linear potential}

The relevant canonical transformations involve an area-preserving shear of the 
phase plane coupled with a shift of the origin: 
\[\left[\begin{array}{c}Q\\ P\end{array}\right]=
\left(\begin{array}{cc}1& -2\nu\\ 0& 1\end{array}\right)
\left[\begin{array}{c}q\\ p\end{array}\right]+2m\lambda\nu
\left[\begin{array}{c}-\nu \\ 1\end{array}\right]\equiv
{\cal T}_\nu\left[\begin{array}{c}q\\ p\end{array}\right],\]
where we have introduced the parameter $\nu\equiv 2/\mu$ which makes the 
algebra of these transformations more transparent. (The factor of 2 in the 
definition of $\nu$ is a matter of convenience.) Because of the shift in 
origin, it is perhaps not obvious that this class of geometrical operations 
should form a group, but, in fact, the composition 
\begin{equation}\label{eq:lpcomp}
{\cal T}_{\nu_1}{\cal T}_{\nu_2}={\cal T}_{\nu_1+\nu_2}
\end{equation}
so that the transformations ${\cal T}_\nu$ are a representation of an  
abelian affine group. 

Let $F_\nu(q,Q)$ denote the generating function of these canonical 
transformations when $\nu$ and not $\mu$ is adopted as the free parameter, 
i.e.\ 
\[F_\nu(q,Q)=-m\lambda\nu(q+Q)+\frac{1}{4\nu}(q-Q)^2,\]
and let $\psi_E(q)$ ($E$ real) denote the eigenfunction of 
\[\widehat{h}_{\rm linear}=
-\frac{\hbar^2}{2m}\frac{\partial^2\ }{\partial q^2} + \lambda q\]
with eigenenergy $E$. Identifying $F_1$, $F_2$ and $F_c$ in 
Eq.~(\ref{eq:qgfconv}) with $F_{\nu_1}$, $F_{\nu_2}$ and $F_{\nu_1+\nu_2}$, 
respectively, we find that the reciprocals $N_E(\nu)$ of the eigenvalues in 
the integral equation involving the eigenfunction $\psi_E(q)$ must satisfy
\begin{equation}\label{eq:Nlinfun}
N_E(\nu_1+\nu_2)=\sqrt{4\pi\hbar i}
\sqrt{\frac{\nu_1\nu_2}{\nu_1+\nu_2}}
e^{-(im^2\lambda^2/\hbar)\nu_1\nu_2(\nu_1+\nu_2)} N_E(\nu_1)N_E(\nu_2).
\end{equation}
Equation (\ref{eq:Nlinfun}) determines $N_E(\nu)$ to be of the form
\[N_E(\nu)=\frac{1}{\sqrt{4\pi\hbar i\nu}}
e^{(i/\hbar)(c_E\nu-m^2\lambda^2\nu^3/3)}.\]
The $E$ dependence of the coefficient $c_E$ can be pinned down by using the 
relation of eigenfunctions
of non-zero energy $\psi_E(q)$ to the zero energy eigenfunction $\psi_0(q)$: 
\begin{equation}\label{eq:Escaling}
\psi_E(q)=\eta_E\psi_0(q-E/\lambda),
\end{equation}
where $\eta_E$ is a constant of modulus unity which we assume below is absorbed 
into the definition of $\psi_E(q)$ with an appropriate choice of the phase. 

Using Eq.~(\ref{eq:Escaling}) to substitute for $\psi_E(Q)$ and $\psi_E(q)$ in 
the integral equation
\begin{equation}\label{eq:linint}
\psi_E(q)=N_E(\nu)\int e^{(i/\hbar) F_\nu(q,Q)}\psi_E(Q){\rm d}Q,
\end{equation}
we find, after the change of variable $Q\rightarrow Q^\prime=Q-E/\lambda$,
\[\psi_0(q-E/\lambda)=N_E(\nu)e^{-i2mE\nu/\hbar}
\int e^{(i/\hbar) F_\nu(q-E/\lambda,Q^\prime)}\psi_0(Q^\prime){\rm d}Q^\prime
=\frac{N_E(\nu)}{N_0(\nu)}e^{-i2mE\nu/\hbar}\psi_0(q-E/\lambda),\]
where to obtain the last equality we have invoked Eq.~(\ref{eq:linint}) again. 
The choice $c_E=2mE$ is indicated.

An independent check of these results is given by working with the momentum 
space equivalent of Eq.~(\ref{eq:linint}):
\begin{equation}\label{eq:mosplint}
\widetilde{\psi}_E(p)=N_E(\nu)\int K(p,P)\widetilde{\psi}_E(P)
{\rm d}P,
\end{equation}
where the kernel
\[K(p,P)\equiv\int e^{-ipq/\hbar}e^{(i/\hbar)F_\nu(q,Q)}
e^{+iPQ/\hbar} \frac{{\rm d}Q{\rm d}q}{2\pi\hbar}=
\sqrt{4\pi\hbar i\nu}
e^{-i[\nu/(4\hbar)](p+P)^2}\delta(P-p-2m\lambda\nu)\]
and the Fourier transform
\[\widetilde{\psi}_E(p)\equiv\frac{1}{\sqrt{2\pi\hbar}}\int e^{-ipq/\hbar}
\psi_E(q){\rm d}q= e^{-iEp/(\hbar\lambda)}\widetilde{\psi}_0(p).\]
The delta function in $K(p,P)$ enforces the relation between the
momenta in the canonical transformation generated by $F_\nu(q,Q)$
and reduces Eq.~(\ref{eq:mosplint}) to the algebraic equation
\[\widetilde{\psi}_0(p)e^{-ip^3/(6m\lambda\hbar)}=N_E(\nu)\sqrt{4\pi\hbar i\nu}
e^{-(i/\hbar)(2mE\nu-m^2\lambda^2\nu^3/3)}
\widetilde{\psi}_0(p+2m\lambda\nu)e^{-i(p+2m\lambda\nu)^3/(6m\lambda
\hbar)}.\]
Since $\widetilde{\psi}_0(p)=Ce^{+ip^3/(6m\lambda\hbar)}$, where $C$ is a 
normalization constant, we recover the results above for $N_E(\nu)$.

The zero energy eigenfunction $\psi_0(q)$ [and hence all the other 
eigenfunctions $\psi_E(q)$] is related to the Airy function 
$\mbox{Ai}(x)$~\cite{AS64}:
\[\psi_0(q)=\frac{\gamma}{\sqrt{\lambda}}\mbox{Ai}(\gamma q),\] 
where $\gamma\equiv(2m\lambda/\hbar^2)^{1/3}$. [For the sake of definiteness, 
the normalization is fixed so that $\left\langle\psi_E|\psi_E^\prime\right
\rangle=\delta(E-E^\prime)$.]
Thus, in terms of the Airy function $\mbox{Ai}(x)$, our integral equation 
Eq.~(\ref{eq:linint}) amounts to the relation
\[\mbox{Ai}(x)=\frac{1}{\sqrt{4\pi is}}e^{-is^3/12}\int
e^{i[-s(x+X)/2+(x-X)^2/(4s)]}\mbox{Ai}(X){\rm d}ýX,\]
where we have set $x=\gamma q$, $X=\gamma Q$ and $s=\hbar\gamma^2\nu$.

\section{Integral equations for the other potentials}\label{sc:op}

In Sec.~\ref{sc:ihf} we found that, for potentials ${\cal V}(x)$ for 
which ${\cal V}^{(3)}\not= 0$, non-trivial form-preserving correction-free 
generating functions ${\cal F}_\mu(q,Q)$ only exist if
${\cal V}^{\prime\prime}(x)=\rho{\cal V}(x)$
($\rho$ a constant). A related implication is that the dependence of 
these generating functions on $q$ and $Q$ must be such that
\begin{equation}\label{eq:qQpds}
\frac{\partial^2{\cal F}_\mu}{\partial q^2}=\rho{\cal F}_\mu=
\frac{\partial^2{\cal F}_\mu}{\partial Q^2}.
\end{equation}
The dependence on $\mu$ is such that
\[\left(\mu\frac{\partial\ }{\partial\mu}\right)^2{\cal F}_\mu={\cal F}_\mu,\]
which, setting $\mu=\mu(z)\equiv\mu_0e^{\sqrt{\rho}z}$, becomes
\begin{equation}\label{eq:zpds}
\frac{\partial^2\ }{\partial z^2}{\cal F}_{\mu(z)}=\rho{\cal F}_{\mu(z)}.
\end{equation}
The similarity of Eqs.~(\ref{eq:qQpds}) and (\ref{eq:zpds}) suggests that it 
should be possible to treat $z$ in ${\cal F}_{\mu(z)}(q,Q)$ in formally the 
same way as the generalized co-ordinates $q$ and $Q$. In fact, we find that, 
with appropriate choices of $\mu_0$ (which are listed 
in Table~\ref{tb:dual}), 
\begin{equation}\label{eq:dual}
{\cal F}_{\mu(z)}(q,Q)={\cal F}_{\mu(q)}(z,Q)={\cal F}_{\mu(Q)}(q,z)
\end{equation}
confirming that the roles of $z$ and $q$ (or $z$ and $Q$) may be interchanged.

Used in conjunction with the integral equation Eq.~(\ref{eq:bie}),
Eq.~(\ref{eq:dual}) implies that
\[\frac{\psi_\alpha(q)}{{\cal N}_\alpha\left(\mu(z)\right)}
=\int e^{(i/\hbar){\cal F}_{\mu(z)}(q,Q)}\psi_\alpha(Q){\rm d}Q=
\int e^{(i/\hbar){\cal F}_{\mu(q)}(z,Q)}\psi_\alpha(Q){\rm d}Q=
\frac{\psi_\alpha(z)}{{\cal N}_\alpha\left(\mu(q)\right)}.\]
Thus, the  reciprocals ${\cal N}_\alpha(\mu)$ of eigenvalues are given to 
within a constant $C_\alpha$ by
\begin{equation}\label{eq:nrm}
{\cal N}_\alpha(\mu)=\frac{C_\alpha}{\psi_\alpha(z)},
\end{equation}
where $z=z(\mu)\equiv\ln(\mu/\mu_0)^{1/\sqrt{\rho}}$.
The constant $C_\alpha$ can be fixed by the requirement that
\begin{equation}\label{eq:largemu}
\lim_{\mu\rightarrow\infty}{\cal N}_\alpha(\mu)e^{(i/\hbar){\cal F}_\mu(q,Q)}=
\delta(q-Q),
\end{equation}
reflecting the fact that, in the limit $\mu\longrightarrow\infty$, we recover 
the identity transformation from ${\cal F}_\mu(q,Q)$ (cf.\ the end of
Sec.~\ref{sc:ihf}). 
                                                                                
To establish the limit on the lefthand-side of Eq.~(\ref{eq:largemu}), we 
consider the integral
\begin{equation}\label{eq:spint}
\int e^{(i/\hbar){\cal F}_\mu(q,Q)}\Phi(Q){\rm d}Q, 
\end{equation}
where $\Phi(Q)$ is a suitable test function, and apply the method of stationary 
phase to extract the leading contribution to the integral when $\mu\gg 1$.
 
By way of illustration, we now discuss the cases of the exponential 
and sinusoidal potentials in more detail.

\subsection{The exponential potential}

The hamiltonian
\[\widehat{h}_{\rm exponential}=-\frac{\hbar^2}{2m}\frac{\partial^2\ }{
\partial q^2}+\frac{\lambda}{2a} e^{2aq}\]
has eigenfunctions of energy $E_k=\hbar^2k^2/(2m)$ given up to a normalization 
constant by
\begin{equation}\label{eq:exppsi}
\psi_k(q)\propto K_{i(k/a)}\left(\sqrt{m\lambda a} e^{aq}/[\hbar a^2]\right),
\end{equation} 
where $K_\sigma$ denotes a modified Bessel function of order $\sigma$ 
(section 9.6 in \cite{AS64}). Substitution of Eq.~(\ref{eq:exppsi}) into 
Eq.~(\ref{eq:nrm}) yields for the reciprocals of eigenvalues
\begin{equation}\label{eq:nrexp}
{\cal N}_k(\mu)=C_k\left/K_{i(k/a)}\left(\mu/[4i\hbar a^2]\right)\right.,
\end{equation} 
where the constant $C_k$ has still to be determined by consideration of the 
$\mu\longrightarrow\infty$ limit.

Asymptotic analysis for $\mu\gg 1$ via the method of stationary phase implies 
that, to leading order, the integral
\[\int e^{(i/\hbar){\cal F}_\mu(q,Q)}\Phi(Q){\rm d}Q\sim 
\sqrt{2\pi}\sqrt{\frac{4\hbar i}{\mu}}e^{-\mu/(4\hbar ia^2)}\Phi(q).\]
Thus, Eq.~(\ref{eq:largemu}) is satisfied provided
\begin{equation}
\lim_{\mu\rightarrow\infty}\sqrt{\frac{4\hbar ia^2}{\mu}}e^{-\mu/(4\hbar 
i a^2)}{\cal N}_k(\mu)=\frac{a}{\sqrt{2\pi}},
\end{equation}
which, on the substitution of Eq.~(\ref{eq:nrexp}) and use of the leading 
term in the asymptotic expansion of $K_\sigma(x)$ for $x\gg 1$,
reduces to the requirement that $C_k=a/2$.

In terms of the variables $y\equiv\sqrt{m\lambda a}e^{aq}/(\hbar a^2)$,
$Y\equiv\sqrt{m\lambda a}e^{aQ}/(\hbar a^2)$, $p=ik/a$ and $w\equiv\mu/
(4\hbar ia^2)$, our integral equation reads
\[2K_p(w)K_p(y)=
\int\limits_0^\infty e^{-[yY/w+w(y/Y+Y/y)]/2}K_p(Y)\frac{{\rm d}Y}{Y},\]
which coincides formally with Eq.~6.653.2 in~\cite{GR94} (after 
the change of integration variable $Y\longrightarrow x\equiv wy/Y$).

\subsection{The sinusoidal potential}

We confine our attention to the denumerable set of eigenfunctions 
$\psi_s(q)$ ($s=0,\pm 1,\pm 2,\ldots$) of
\[\widehat{h}_{\rm sinusoidal}=-\frac{\hbar^2}{2m}\frac{\partial^2\ }{
\partial q^2}+\frac{\lambda}{4a^2}\cos 2aq\]
which are related to the Mathieu functions $ce_r$ and $se_r$: 
\begin{equation}\label{eq:psim}
\psi_s(q)=\left\{\begin{array}{c@{\qquad}l}{\cal C}_s\,ce_s(aq,\delta)& 
s=0,1,2,\ldots \\[1pc]
{\cal C}_s\,se_{|s|}(aq,\delta) & s=-1,-2,\dots,\end{array}\right.
\end{equation}
where the dimensionless rescaled strength of the potential 
$\delta\equiv m\lambda/(4
\hbar^2a^4)$ and ${\cal C}_s$ denotes a normalization constant.

The choice of the range of integration in the integral equation for the 
$\psi_s$'s
\[\psi_s(q)=N_s(\mu)\int e^{i{\cal F}_\mu(q,Q)}\psi_s(Q){\rm d}Q\]
is dictated by the consideration that the 
bilinear concomitant in Eq.~(\ref{eq:bilcom}) vanishes. 
This can be achieved by exploiting the periodicity of 
the generating function ${\cal F}_\mu(q,Q)$ (period $2\pi/a$
in $q$ or $Q$) and the eigenfunctions $\psi_s(q)$ (period $\pi/a$ for $s$ 
even and $2\pi/a$ for $s$ odd). Thus, we take the range of integration
to be over one period of the generating function from $0$ to $+2\pi/a$.
(Use of these non-symmetric limits facilitates the comparison with the 
integral equations tabulated in ch.~20 of \cite{AS64}.) 

Substituting for the wavefunction in Eq.~(\ref{eq:nrm}) using 
Eq.~(\ref{eq:psim}) and invoking
the proportionality of $ce_r(-ix,\eta)$ and $se_r(-ix,\eta)$ to the 
modified Mathieu functions $Mc^{(1)}_r(x,\eta)$ and $Ms^{(1)}_r(x,\eta)$, 
respectively, we find that
\begin{equation}\label{eq:Mathieunrm}
N_s(\mu)=\frac{C_s}{M^{(1)}_s\left(\ln[\mu/\sqrt{4m\lambda}],\delta
\right)},
\end{equation}
where $C_s$ is independent of $\mu$ and 
$M_s^{(1)}(x,\eta)$ denotes the modified Mathieu function $Mc^{(1)}_s
(x,\eta)$ for $s\ge 0$ and the modified Mathieu 
function $Ms^{(1)}_{|s|}(x,\eta)$ for $s<0$.  

In the asymptotic analysis of the integral in Eq.~(\ref{eq:spint}) for large 
$\mu$, we encounter in the present case two points of stationary phase: 
$Q=Q_1=q+O(\mu^{-2})$ and $Q=Q_2=q-\sigma_q\pi/a+O(\mu^{-2})$, where 
$\sigma_q$ denotes the sign of $q-\pi/a$. To leading order, the asymptotic 
expansion reads
\begin{eqnarray}
\lefteqn{\int\limits_0^{2\pi/a}e^{(i/\hbar){\cal F}_\mu(q,Q)}\Phi(Q){\rm d}Q} 
\nonumber\\ 
& &\mbox{}\sim\sqrt{\frac{8\pi\hbar}{\mu}}\left[e^{-i\mu/(4\hbar a^2)+i\pi/4}
\Phi(q)
+e^{+i\mu/(4\hbar a^2)-i\pi/4}\Phi(q-\sigma_q\pi/a)\right].\label{eq:sinsp}
\end{eqnarray}
To proceed, it is necessary to recognise that the choice of the appropriate
space of test functions $\Phi(q)$ depends on the properties of the 
eigenfunctions $\{\psi_s(q)\}$ under consideration. Accordingly, $\Phi(q)$ is 
drawn from either of two spaces ${\cal S}_p$ ($p=0,1$): a space ${\cal S}_0$ 
of periodic functions of period $\pi/a$ appropriate to the eigenfunctions 
$\psi_s$ for even $s$, and: a space ${\cal S}_1$ of periodic functions of 
period $2\pi/a$ appropriate to the eigenfunctions $\psi_s$ for odd $s$.
Furthermore, paralleling the property that $\psi_s(q-\sigma_q\pi/a)=
(-1)^s
\psi_s(q)$, we must require that test functions drawn from ${\cal S}_1$ are 
such that $\Phi(q-\sigma_q\pi/a)=-\Phi(q)$. Thus, for test functions 
drawn from ${\cal S}_p$, Eq.~(\ref{eq:sinsp}) becomes
\begin{equation}\label{eq:sinsift}
\int\limits_0^{2\pi/a}e^{(i/\hbar){\cal F}_\mu(q,Q)}\Phi(Q){\rm d}Q
\sim 2i^{-p}\sqrt{\frac{8\pi\hbar}{\mu}}\cos[\mu/(4\hbar a^2)-(p+1/2)\pi/2]
\Phi(q)
\end{equation}
so that, for a suitable choice of $N_s(\mu)$, $\lim_{\mu\longrightarrow\infty}
N_s(\mu)e^{(i/\hbar){\cal F}_\mu(q,Q)}$ can have the sifting property expected.

To leading order,
the asymptotic expansion of $M_s^{(1)}(\ln[\mu/\sqrt{4m\lambda}],\delta)$ in 
the limit of large $\mu$ ($>0$) is [from the real part of Eq.~(20.9.1) in 
\cite{AS64}]
\begin{equation}\label{eq:Mathieuasymptotic}
M_s^{(1)}\left(\ln[\mu/\sqrt{4m\lambda}],\delta\right)\sim 
\frac{i^{|s|-p}}{\sqrt{\pi}}\sqrt{\frac{8\hbar a^2}{\mu}}
\cos[\mu/(4\hbar a^2)-(p+1/2)\pi/2],
\end{equation}
where $p=0$ (1) for $s$ even (odd). [Despite appearances, the righthand-side of 
Eq.~(\ref{eq:Mathieuasymptotic}) is real-valued consistent with the reality of 
$M_s^{(1)}(x,\eta)$ for real-valued arguments $x$ and $\eta$.]
Combining Eqs.~(\ref{eq:Mathieunrm}),
(\ref{eq:sinsift}) and (\ref{eq:Mathieuasymptotic}) in Eq.~(\ref{eq:largemu}),
we conclude that $C_s=i^{|s|}a/(2\pi)$.

Introducing $\zeta\equiv\ln[\mu/\sqrt{4m\lambda}]$ and the variables 
$u\equiv q/a$ and $U\equiv Q/a$, the integral equation for the $\psi_s$'s reads
\[\psi_s(u/a)=\frac{i^{|s|}}{2\pi}\frac{1}{M^{(1)}_s(\zeta,\delta)}
\int\limits_0^{2\pi}e^{-2i\sqrt{\delta}(\cosh\zeta\cos u\cos U+\sinh\zeta
\sin u\sin U)}\psi_s(U/a){\rm d}U,\]
which for $\zeta$ real is tantamount to the complex conjugate of Eqs.~(20.7.34) 
and (20.7.35) in~\cite{AS64}.

\section{Discussion}\label{sc:dis}

In this paper, we have explored the implications for quantum mechanical systems
of the existence of non-trivial canonical transformations which leave the form  
of the Hamilton function of the corresponding classical system invariant. 
Such dynamical symmetries of a classical system manifest themselves in a linear 
homogeneous integral equation of the kind in Eq.~(\ref{eq:bie})
for the stationary state eigenfunctions of the quantum 
mechanical system. We have seen that the quantum canonical transform introduced 
in~\cite{Gh87} is ideally suited to the purpose of tying down 
the features of these integral equations. 

We have not exhausted the full range of integral equations which can be 
constructed by invoking the quantum canonical transform. For the families of 
canonical transformations which we have 
considered which do not constitute Abelian groups, there is a (possibly 
infinite) sequence of integral equations corresponding to the repeated 
composition of these transformations. Consideration of the composition of a 
transformation to angle-action variables with its inverse could give rise to 
still
more integral equations. However, the most interesting line of 
further investigation in our opinion is the extension of the work in this
paper to quantum field theory. 

There are some almost immediate parallels of our results for theories of a
scalar field $\varphi(\sigma,\tau)$ in 1+1 dimensions ($\sigma$ denotes the 
spatial dimension and $\tau$ the time in natural 
units such that $\hbar=1=c$). This is in part a consequence of the fact that
the (first quantized) Hamiltonian functionals $H[\varphi,\pi]$ 
for these theories are not too dissimilar in form from the Hamilton functions 
$\cal H$ we have considered:
\[H[\varphi,\pi]={\textstyle\frac{1}{2}}\int\left[\pi^2+\left(\left.\partial
\varphi\right/\partial\sigma\right)^2
\right]{\rm d}\sigma+\int V(\varphi){\rm d}\sigma,\]
where $\pi$ is the field momentum conjugate to $\varphi$ and the ``potential 
density'' $V$ describes the self-coupling of $\varphi$. If we consider
canonical transformations $\varphi,\pi\longrightarrow\Phi,\Pi$ induced by
generating functionals of the form [$\varphi_\pm\equiv(\varphi\pm\Phi)/2$]
\[F[\varphi,\Phi]=\int\varphi\frac{\partial\Phi}{\partial\sigma}{\rm d}\sigma+
\int\left[F_+(\varphi_+)+F_-(\varphi_-)\right]{\rm d}\sigma,\]
then we find that a sufficient condition for the transformed Hamilton 
functional to be of the same form as the original Hamilton functional is that
\begin{equation}\label{eq:fbasic}
\frac{1}{2}\frac{\partial F_+}{\partial\varphi_+}
\frac{\partial F_-}{\partial\varphi_-}=
V(\varphi_+-\varphi_-)-V(\varphi_++\varphi_-)
\end{equation}
provided $F_\pm$ either vanish or are periodic at the endpoints of 
the integration over $\sigma$. Apart from an inessential factor of $m$, 
Eq.~(\ref{eq:fbasic}) is formally identical to the condition
established in Sec.~\ref{sc:ihf} for the invariance of Hamilton functions under
canonical transformations [Eq.~(\ref{eq:basic})] with $V(\varphi)$ replacing 
the potential ${\cal V}(q)$ and $\varphi_\pm(\sigma,\tau)$ the combinations 
$q_\pm=(q\pm Q)/2$. In the Schr{\"o}dinger representation~\cite{Ha92} of the 
corresponding (second) quantized field theories, we have, putting aside the 
issue of renormalization, a class of integral equations for the wavefunctionals 
$\Psi_\alpha$ of the form
\[\Psi_\alpha[\varphi]={\cal N}_\alpha[\mu]\int e^{iF_\mu[\varphi,\Phi]}
\Psi_\alpha[\Phi]{\cal D}\Phi,\]
where, in terms of the generating functions ${\cal F}_\mu$ listed in 
Table~\ref{tb:potentials},
\[F_\mu[\varphi,\Phi]=\int\varphi\frac{\partial\Phi}{\partial\sigma}
{\rm d}\sigma+\int{\cal F}_\mu(\varphi,\Phi){\rm d}\sigma\]
and ${\cal N}_\alpha$ is a functional of $\mu$ which is now a function of 
$\sigma$ and $\tau$.

% Tables

\begin{table}
\caption{Standard forms of potentials $\cal V$ and the corresponding (classical) 
generating functions ${\cal F}_\mu$}
\label{tb:potentials}
\begin{tabular}{*{3}{l}}
             & ${\cal V}(q)$            & ${\cal F}_\mu(q,Q)$    \\ \hline
             &                          &                  \\
Quadratic    &$\displaystyle\frac{1}{2}\lambda q^2$  &
 $-\displaystyle\frac{m\lambda}{2\mu}(q+Q)^2+\frac{\mu}{8}(q-Q)^2$\\[1pc]
Sinusoidal   &$\displaystyle\frac{\lambda}{4a^2}\cos 2aq$ & 
 $-\displaystyle\frac{m\lambda}{\mu a^2}\cos a(q+Q)-\frac{\mu}{4a^2}\cos 
  a(q-Q)$\\[1pc]
Even Hyperbolic &$\displaystyle\frac{\lambda}{4a^2}\cosh 2aq$ &
 $-\displaystyle\frac{m\lambda}{\mu a^2}\cosh a(q+Q)+\frac{\mu}{4a^2}\cosh
  a(q-Q)$\\[1pc]
Linear       &$\lambda q$               &
 $-\displaystyle\frac{2m\lambda}{\mu}(q+Q)+\frac{\mu}{8}(q-Q)^2$\\[1pc]
Exponential  &$\displaystyle\frac{\lambda}{2a}e^{2aq}$ &
 $-\displaystyle\frac{2m\lambda}{\mu a}e^{a(q+Q)}+\frac{\mu}{4a^2}\cosh 
  a(q-Q)$\\[1pc]
Odd Hyperbolic &$\displaystyle\frac{\lambda}{2a}\sinh 2aq$ &
 $-\displaystyle\frac{2m\lambda}{\mu a}\sinh a(q+Q)+\frac{\mu}{4a^2}\cosh 
  a(q-Q)$\\
             &                          &                  \\
\end{tabular}
\end{table}

\begin{table}
\caption{Parameters of the transformation
$z(\mu)=\rho^{-1/2}\ln(\mu/\mu_0)$}
\label{tb:dual}
\begin{tabular}{*{3}{l}}
${\cal V}(q)$   &     $\mu_0$            & $\sqrt{\rho}$   \\ \hline
                &                        &                 \\
Sinusoidal      & $2\sqrt{m\lambda}$    & $ia$            \\[1pc]
Even hyperbolic & $2\sqrt{m\lambda}i$    & $a$             \\[1pc]
Exponential     & $4\sqrt{m\lambda a}i$  & $a$             \\[1pc]
Odd hyperbolic  & $2\sqrt{2m\lambda a}i$ & $a$             \\
                &                        &                 \\ 
\end{tabular}
\end{table}

\end{document}